\documentstyle[12pt,oldlfont,epsf,rotating]{article}
\newcommand{\n}{\hspace*{-2.5mm}}
\newcommand{\sla}{\hspace*{.5mm}\slash\hspace*{-2.3mm}}
\newcommand{\isla}{\hspace*{.5mm}\slash\hspace*{-1.7mm}}
\newcommand{\gsim}{\;\rlap{\lower 3.5 pt \hbox{$\mathchar \sim$}} \raise 1pt
 \hbox {$>$}\;}
\newcommand{\lsim}{\;\rlap{\lower 3.5 pt \hbox{$\mathchar \sim$}} \raise 1pt
 \hbox {$<$}\;}

\catcode`@=11
\newcount\@tempcntc
\def\@citex[#1]#2{\if@filesw\immediate\write\@auxout{\string\citation{#2}}\fi
  \@tempcnta\z@\@tempcntb\m@ne\def\@citea{}\@cite{\@for\@citeb:=#2\do
    {\@ifundefined
       {b@\@citeb}{\@citeo\@tempcntb\m@ne\@citea\def\@citea{,}{\bf ?}\@warning
       {Citation `\@citeb' on page \thepage \space undefined}}%
    {\setbox\z@\hbox{\global\@tempcntc0\csname b@\@citeb\endcsname\relax}%
     \ifnum\@tempcntc=\z@ \@citeo\@tempcntb\m@ne
       \@citea\def\@citea{,}\hbox{\csname b@\@citeb\endcsname}%
     \else
      \advance\@tempcntb\@ne
      \ifnum\@tempcntb=\@tempcntc
      \else\advance\@tempcntb\m@ne\@citeo
      \@tempcnta\@tempcntc\@tempcntb\@tempcntc\fi\fi}}\@citeo}{#1}}
\def\@citeo{\ifnum\@tempcnta>\@tempcntb\else\@citea\def\@citea{,}%
  \ifnum\@tempcnta=\@tempcntb\the\@tempcnta\else
   {\advance\@tempcnta\@ne\ifnum\@tempcnta=\@tempcntb \else \def\@citea{--}\fi
    \advance\@tempcnta\m@ne\the\@tempcnta\@citea\the\@tempcntb}\fi\fi}
\catcode`@=12
\begin{document}
\title{\vskip-3cm{\baselineskip14pt
\centerline{\normalsize DESY 94-102\hfill ISSN 0418-9833}
\centerline{\normalsize October 1994\hfill hep-ph/9410319}
}
\vskip1.5cm
Two-Loop ${\cal O}(\alpha_sG_Fm_t^2)$ Correction to the $H\to b\bar b$
Decay Rate}
\author{Bernd A. Kniehl\\
Max-Planck-Institut f\"ur Physik, Werner-Heisenberg-Institut\\
F\"ohringer Ring 6, 80805 Munich, Germany\\ \\
Michael Spira\\
Theory Group, DESY\\
Notkestra\ss e 85, 22603 Hamburg, Germany}
\date{}
\maketitle
\begin{abstract}
We present the two-loop ${\cal O}(\alpha_sG_Fm_t^2)$ correction to the
$b\bar b$ decay rate of the Stan\-dard-Model Higgs boson, assuming that
the $t$ quark is much heavier than the Higgs boson.
Apart from the universal correction connected with the renormalizations
of the wave function and the vacuum expectation value of the Higgs field,
this involves vertex corrections specific to the presence of beauty in the
final state.
We calculate the latter by means of a low-energy theorem.
All would-be mass singularities related to the $b$ quark can be absorbed
into the running Higgs-bottom Yukawa coupling.
It turns out that the total ${\cal O}(\alpha_sG_Fm_t^2)$ correction screens
the leading high-$m_t$ behaviour of the one-loop result by 71\% to 75\% for
$M_H$ between 60~GeV and $2M_W$.
\end{abstract}

\section{Introduction}

One of the great puzzles of elementary particle physics today
is whether nature makes use of the Higgs mechanism of
spontaneous symmetry breaking to generate the observed particle masses.
The Higgs boson, $H$, is the missing link sought to verify
this concept in the Standard Model.
The failure of experiments at LEP~1 and SLC to detect the decay
$Z\rightarrow f\bar f H$ has ruled out the mass range $M_H\le63.8$~GeV at
the 95\% confidence level \cite{sch}.
At the other extreme, unitarity arguments in intermediate-boson scattering
at high energies \cite{dic} and considerations concerning the range of
validity of perturbation theory \cite{vel} establish an upper bound on
$M_H$ at $\left(8\pi\sqrt2/3G_F\right)^{1/2}\approx1$~TeV in a weakly
interacting Standard Model.

A Higgs boson with $M_H\lsim135$~GeV decays dominantly to $b\bar b$ pairs
\cite{bak}.
This decay mode will be of prime importance for Higgs-boson searches at
LEP~2 \cite{gkw}, the Tevatron \cite{mar}---or a possible 4-TeV upgrade
thereof \cite{gun}---, and the next $e^+e^-$ linear collider \cite{imh}.
Techniques for the measurement of the $H\to b\bar b$ branching fraction at a
$\sqrt s=500$~GeV $e^+e^-$ linear collider have been elaborated in
Ref.~\cite{hil}.

The present knowledge of quantum corrections to the $H\to b\bar b$ decay
rate has been reviewed very recently in Ref.~\cite{bak}.
The QCD corrections are most significant numerically.
In ${\cal O}(\alpha_s)$, their full $m_b$ dependence is known \cite{bra}.
In ${\cal O}(\alpha_s^2)$, the first \cite{gor} and second \cite{sur} terms
of the expansion in $m_b^2/M_H^2$ have been found.
The large logarithmic corrections of the forms $\alpha_s^n\ln^m(M_H^2/m_b^2)$
and $\alpha_s^nm_b^2/M_H^2\ln^m(M_H^2/m_b^2)$, with $n\ge m$ and $n,m=1,2$,
which are present when the on-shell scheme of quark mass renormalization
is employed, may be succinctly absorbed into the running $b$-quark mass
of the $\overline{\rm MS}$ scheme evaluated at scale $M_H$.
In this way, these logarithms are resummed to all orders and the perturbation
series converges more rapidly.
The residual terms are perturbatively small.
The theoretical uncertainty related to the lack of knowledge of the
nonlogarithmic ${\cal O}(\alpha_s^3)$ term is likely to be
irrelevant for practical purposes \cite{kat}.
The QCD correction relative to the Born approximation implemented with
the pole mass ranges between $-53\%$ and $-63\%$ for $M_H$ between 60~GeV
and $2M_W$ \cite{bak}.

The full one-loop electroweak corrections to the $H\to b\bar b$ decay width
are well established \cite{fle,hff}.
They consist of an electromagnetic and a weak part, which are separately
finite and gauge independent.
The electromagnetic part emerges from the ${\cal O}(\alpha_s)$ correction
in the on-shell scheme \cite{bra} by substituting $\alpha Q_b^2$ for
$\alpha_sC_F$, where $Q_b=-1/3$ and $C_F=\left(N_c^2-1\right)/(2N_c)$,
with $N_c=3$.
For $M_H\ll2M_W$, the weak part is well approximated by \cite{hff}
\begin{equation}
\label{weak}
\Delta_{\rm weak}={G_F\over8\pi^2\sqrt2}\left\{m_t^2
+M_W^2\left({3\over s_w^2}\ln c_w^2-5\right)
+M_Z^2\left[{1\over2}-3\left(1-4s_w^2|Q_b|\right)^2\right]\right\},
\end{equation}
where $c_w^2=1-s_w^2=M_W^2/M_Z^2$.
Here it is understood that the Born result is expressed in terms of
the Fermi constant \cite{res},
\begin{equation}
\label{born}
\Gamma_0\left(H\to b\bar b\right)={N_cG_FM_Hm_b^2\over4\pi\sqrt2}
\left(1-{4m_b^2\over M_H^2}\right)^{3/2}.
\end{equation}

Equation~(\ref{weak}) has been obtained by putting $M_H=m_f=0$
($f\ne t$) in the expression for the full one-loop weak correction \cite{hff}.
It provides a very good approximation up to $M_H\approx70$~GeV,
the relative deviation from the full weak correction being less than 15\%.
In view of evidence for a heavy $t$ quark, with $m_t=(174\pm16)$~GeV
\cite{cdf}, the first term of Eq.~(\ref{weak}) is particularly important.
It consists of a universal part, which contributes to all fermionic
Higgs-boson decays, and a non-universal part, which is specific to
$b\bar b$ production.
The universal part arises from the renormalizations of the wave function
and the vacuum expectation value of the Higgs field, while the non-universal
part is due to the $b\bar bH$ vertex correction and the $b$-quark
wave-function renormalization.
At one loop, there is a large cancellation between the universal and
non-universal parts, their sum being seven times smaller than the universal
part alone.

The gluon correction to the shift in $\Gamma\left(H\to f\bar f\right)$
due to a doublet of novel quarks with arbitrary masses has been evaluated
recently
in Ref.~\cite{uni}.
The QCD correction to the ${\cal O}(G_Fm_t^2)$ universal term
emerges as a special case from this result \cite{hll}.
It screens the one-loop term in such a way that the corrected term is given
by $2\delta_{\rm u}$, where
\begin{equation}
\label{dun}
\delta_{\rm u}=x_t\left[{7\over2}-{3\over2}\left(\zeta(2)+{3\over2}\right)
C_F{\alpha_s\over\pi}\right],
\end{equation}
with $x_t=\left(G_Fm_t^2/8\pi^2\sqrt2\right)$.
This correction constitutes the full ${\cal O}(\alpha_sG_Fm_t^2)$ contribution
to all fermionic decay widths of a Higgs boson with $M_H<2m_t$, except for the
$b\bar b$ channel.
In the latter case, one needs to include the gluon correction to the
${\cal O}(G_Fm_t^2)$ non-universal term.
The evaluation of this correction is the subject of this article.
For this purpose, we shall take advantage of a low-energy theorem
\cite{ell,vai,daw}.
Our final result disagrees with a recent analysis \cite{kwi}.
We shall pin down the error in Ref.~\cite{kwi}.

This paper is organized as follows.
In Sect.~2, we shall describe the low-energy theorem,
which we shall apply in Sect.~3 to derive effective Lagrangians for the
$b\bar bH$ and $b\bar bgH$ interactions to ${\cal O}(\alpha_sG_Fm_t^2)$.
In Sect.~4, we shall use these effective Lagrangians to evaluate the
${\cal O}(\alpha_sG_Fm_t^2)$ correction to $\Gamma\left(H\to b\bar b\right)$.
We also present a master formula for $\Gamma\left(H\to b\bar b\right)$,
which includes all known corrections.
Our conclusions are summarized in Sect.~5.

\section{Low-energy theorem}

Low-energy theorems for Higgs-boson interactions have been studied in great
detail in the literature \cite{ell,vai,daw}.
These theorems relate the amplitudes of two processes which differ by the
insertion of an external Higgs-boson line carrying zero momentum.
They provide a convenient tool for estimating the properties of a Higgs boson
that is light as compared to the loop particles.
For instance, they have been applied to derive low-$M_H$ effective Lagrangians
for the $\gamma\gamma H$ and $ggH$ interactions at one \cite{vai} and two loops
\cite{daw}.

These low-energy theorems may be derived by observing that the
interactions of the Higgs boson with the massive particles
in the Standard Model emerge from their mass terms by substituting
$m_i\to m_i(1+H/v)$, where $m_i$ is the mass of the respective particle,
$H$ is the Higgs field, and $v$ is the Higgs vacuum expectation value.
On the other hand, a Higgs boson with zero momentum is represented by a
constant
field, since $i\partial_\mu H=[P_\mu,H]=0$, where $P_\mu$ is the
four-momentum operator.
This immediately implies that a zero-momentum Higgs boson may be attached
to an amplitude, ${\cal M}(A\to B)$, by carrying out the operation
\begin{equation}
\label{let}
\lim_{p_H\to0}{\cal M}(A\to B+H)={1\over v}\sum_i m_i
{\partial\over\partial m_i}{\cal M}(A\to B),
\end{equation}
where $i$ runs over all massive particles involved in the transition
$A\to B$.
Here it is understood that the differential operator does not act
on the $m_i$ appearing in coupling constants, since this would generate
tree-level four-point interactions involving one Higgs-boson line,
which are absent in the Standard Model.
Special care must be exercised when this low-energy theorem is to be applied
beyond leading order.
Then it must be formulated for the bare quantities of the theory.
The renormalization is performed after the left-hand side of Eq.~(\ref{let})
has been evaluated.

\section{Effective Lagrangians}

We now turn to the ${\cal O}(\alpha_sG_Fm_t^2)$ non-universal correction
to $\Gamma\left(H\to b\bar b\right)$.
Prior to performing the actual calculation, we outline the core of the
procedure.
Inspection of the one-loop weak correction to $\Gamma\left(H\to b\bar b\right)$
\cite{hff} reveals that only diagrams involving virtual $t$ quarks and charged
Higgs-Kibble ghosts, $\phi^\pm$, and without direct $\phi^+\phi^-H$ coupling
contribute to the ${\cal O}(G_Fm_t^2)$ term.
Moreover, the masses of the $\phi^\pm$ scalars may be put to zero.
After factoring out the tree-level $b\bar bH$ amplitude, also the $b$ quark
may be treated as massless, so that the $t$ quark is the only massive particle
left in the loops.

The low-energy theorem (\ref{let}) provides an alternative method of deriving
the ${\cal O}(G_Fm_t^2)$ non-universal correction to the $b\bar bH$ coupling,
which requires only the computation of two-point functions.
In fact, we just need to compute the $b$-quark self-energy amplitude induced by
$t$ and $\phi^\pm$ in the same approximation as above [see Fig.~\ref{one}(a)].
The desired result may then be extracted by differentiation with respect
to the bare $b$- and $t$-quark masses and performing their renormalizations in
the resulting expression.
The $b$-quark wave-function renormalization cancels against the corresponding
part of the $b\bar bH$-vertex counterterm.
By including also the ${\cal O}(G_Fm_t^2)$ universal correction, which
originates in the renormalizations of the Higgs-boson wave function and
vacuum expectation value, we can formulate a low-$M_H$ effective Lagrangian
for the $b\bar bH$ interaction valid to ${\cal O}(G_Fm_t^2)$,
where the $t$ quark is integrated out.

These considerations remain valid when one gluon is attached to the quark
line in all possible ways.
Since the gluon can occur as a virtual or a real particle, at first sight,
it seems that one has to deal with both $b\bar bH$ and $b\bar bgH$ effective
Lagrangians.
However, it is easy to see that the $b\bar bgH$ coupling does not receive
a contribution in ${\cal O}(G_Fm_t^2)$.
This may be understood by observing that the one-loop $b\bar bg$ vertex
correction,
from which the $b\bar bgH$ amplitude may be constructed by means of the
low-energy theorem (\ref{let}),
does not develop a term proportional to $m_t^2$ in the high-$m_t$ limit.
The latter point may be inferred from the analogous calculation of the
$b\bar b\gamma$ vertex at one loop \cite{lyn}.
Proceeding along the same lines as above and exploiting knowledge of the
${\cal O}(\alpha_sG_Fm_t^2)$ universal correction [see Eq.~(\ref{dun})],
it is possible to extend the $b\bar bH$ effective Lagrangian to
${\cal O}(\alpha_sG_Fm_t^2)$.

Finally, we may embed this effective Lagrangian in the usual QCD Lagrangian
involving five quark flavours and perform perturbation theory in $\alpha_s$.
In the case of $\Gamma\left(H\to b\bar b\right)$, we shall then recover the
known ${\cal O}(\alpha_s)$ \cite{bra} and ${\cal O}(\alpha_s^2)$ corrections
\cite{gor,sur} with the leading $m_t$-dependent terms being collected in an
overall factor.
Apart from the advantage of being implemented conveniently,
our result will resum automatically reducible higher-order terms.
By including also the electromagnetic and remaining weak corrections,
we shall obtain the complete Standard-Model prediction.

In summary, our original problem reduces to the evaluation of the self-energy
of an on-shell $b$ quark up to ${\cal O}(\alpha_sG_Fm_t^2)$ in the limit of
vanishing $b$-quark and $W$-boson masses.
The relevant one- and two-loop diagrams are depicted in Figs.~\ref{one}(a) and
(b), respectively.
As usual, we shall use dimensional regularization with anticommuting
$\gamma_5$.
Notice that we have not included the reducible two-loop diagrams in
Fig.~\ref{one}(b), since the one-loop gluon-exchange subdiagram vanishes,
due to the absence of a mass scale to carry its dimension.
These diagrams as well as real-gluon emission will come in at a later stage
through the QCD corrections that are to be evaluated from the effective
Lagrangian keeping the $b$-quark mass finite.
This will be done in the next section.

The bare amplitude characterizing the propagation of an on-shell $b$ quark in
the presence of quantum effects has the form
\begin{equation}
{\cal M}^0(b\to b)=\left[m_b^0\left(-1+\Sigma_S^0(p^2)\right)
+\sla p\left(\Sigma_V^0(p^2)+\gamma_5\Sigma_A^0(p^2)\right)
\right]_{\isla p=m_b^0},
\end{equation}
where $S$, $V$, and $A$ label the scalar, vector, and axial-vector components
of the $b$-quark self-energy, respectively, and the superscript 0 marks bare
quantities.
Using the Dirac equation and putting $m_b^0=0$ in the loop amplitudes,
this becomes
\begin{equation}
{\cal M}^0(b\to b)=m_b^0(-1+\Sigma),
\end{equation}
where $\Sigma=\Sigma_V^0(0)+\Sigma_S^0(0)$.
Evaluating the Feynman diagrams in Figs.~\ref{one}(a) and (b), we obtain the
${\cal O}(G_Fm_t^2)$ and ${\cal O}(\alpha_sG_Fm_t^2)$ terms of
$\Sigma=\Sigma_1+\Sigma_2+\cdots$,
\begin{eqnarray}
\Sigma_1&\n=\n&-x_t^0\left({4\pi\mu^2\over(m_t^0)^2}\right)^\epsilon
\Gamma(1+\epsilon)\left({3\over2\epsilon}+{5\over4}+{\cal O}(\epsilon)\right),
\nonumber\\
\Sigma_2&\n=\n&-C_F{\alpha_s\over\pi}x_t^0
\left({4\pi\mu^2\over(m_t^0)^2}\right)^{2\epsilon}\Gamma^2(1+\epsilon)
\left({9\over8\epsilon^2}+{9\over8\epsilon}+{\cal O}(1)\right),
\end{eqnarray}
where $n=4-2\epsilon$ is the dimensionality of space-time, $\mu$ is the
't~Hooft mass, $\Gamma$ is Euler's gamma function, and
$x_t^0=\left(G_F(m_t^0)^2/8\pi^2\sqrt2\right)$.
We recall that the $m_t$-independent QCD corrections do not contribute to the
effective Lagrangian in the limit of vanishing $b$-quark mass.
Equation~(\ref{let}) now tells us that
\begin{eqnarray}
\lim_{p_H\to0}{\cal M}^0(b\to b+H)
&\n=\n&{1\over v_0}\left(m_t^0{\partial\over\partial m_t^0}
+m_b^0{\partial\over\partial m_b^0}\right){\cal M}^0(b\to b)
\nonumber\\
&\n=\n&{m_b^0\over v_0}\left(-1+\Sigma
+m_t^0{\partial\Sigma\over\partial m_t^0}\right).
\end{eqnarray}
Thus, we evaluate
\begin{eqnarray}
m_t^0{\partial\Sigma_1\over\partial m_t^0}&\n=\n&
x_t^0\left({4\pi\mu^2\over(m_t^0)^2}\right)^\epsilon\Gamma(1+\epsilon)
\left(3+{5\over2}\epsilon+{\cal O}(\epsilon^2)\right),
\nonumber\\
m_t^0{\partial\Sigma_2\over\partial m_t^0}&\n=\n&
C_F{\alpha_s\over\pi}x_t^0\left({4\pi\mu^2\over(m_t^0)^2}\right)^{2\epsilon}
\Gamma^2(1+\epsilon)
\left({9\over2\epsilon}+{9\over2}+{\cal O}(\epsilon)\right),
\end{eqnarray}
treating $x_t^0$ as a constant, since it receives its two powers of $m_t^0$
from the $t\bar b\phi^-$ and $b\bar t\phi^+$ couplings.

Next, we carry out the renormalization procedure.
For this end, we substitute $m_q^0=m_q+\delta m_q$ $(q=t,b)$,
where $m_q$ is the on-shell mass and $\delta m_q$ is the appropriately defined
counterterm.
For $q=b$, we have $\delta m_b/m_b=\Sigma$, so that
\begin{equation}
\label{bot}
\lim_{p_H\to0}{\cal M}^0(b\to b+H)
={m_b\over v_0}\left(-1+m_t^0{\partial\Sigma\over\partial m_t^0}\right),
\end{equation}
which is correct through ${\cal O}(\alpha_sG_Fm_t^2)$.
We observe that $m_t^0(\partial\Sigma_1/\partial m_t^0)$ is already finite in
the physical limit, $\epsilon\to0$, as it must because it constitutes the first
term in the series of leading high-$m_t$ non-universal corrections to the
$b\bar bH$ effective coupling.
$m_t^0(\partial\Sigma_2/\partial m_t^0)$ will become finite when we also
renormalize the $t$-quark mass.
To ${\cal O}(\alpha_s)$, we have
\begin{equation}
{\delta m_t\over m_t}=
-C_F{\alpha_s\over\pi}\left({4\pi\mu^2\over m_t^2}\right)^\epsilon
\Gamma(1+\epsilon)\left({3\over4\epsilon}+1+{\cal O}(\epsilon)\right).
\end{equation}
In fact, this yields an ultraviolet-finite result,
\begin{equation}
\label{dnu}
\delta_{\rm nu}\equiv-m_t^0{\partial\Sigma\over\partial m_t^0}
=x_t\left(-3+{3\over4}C_F{\alpha_s\over\pi}\right),
\end{equation}
where $x_t$ is defined below Eq.~(\ref{dun}).
Note that Eq.~(\ref{dnu}) is $\mu$ independent as it should,
since we are working in the on-shell scheme.
Equation~(\ref{dnu}) reproduces the well-known ${\cal O}(G_Fm_t^2)$
non-universal correction to $\Gamma\left(H\to b\bar b\right)$ \cite{hff}
as may be seen by comparing $2(\delta_{\rm u}+\delta_{\rm nu})$ with
Eq.~(\ref{weak}).
Inserting Eq.~(\ref{dnu}) in Eq.~(\ref{bot}), we obtain
\begin{equation}
\lim_{p_H\to0}{\cal M}^0(b\to b+H)=-{m_b\over v_0}(1+\delta_{\rm nu}).
\end{equation}

We are now in the position to write down the low-$M_H$ effective Lagrangian
for the  $b\bar bH$ interaction including the ${\cal O}(G_Fm_t^2)$ and
${\cal O}(\alpha_sG_Fm_t^2)$ non-universal corrections,
\begin{equation}
\label{eff}
{\cal L}_{\rm eff}=-{m_b\over v^0}\overline bbH^0(1+\delta_{\rm nu}).
\end{equation}
Here we have represented the $b$ quarks by their renormalized fields,
anticipating the cancellation of the corresponding wave-function
renormalizations by an appropriate piece in the $b\bar bH$ vertex counterterm
\cite{hff}.
We still need to include the universal corrections.
They enter through the relation $H^0/v^0=(H/v)(1+\delta_{\rm u})$,
where $v=2^{-1/4}G_F^{-1/2}$ and $\delta_{\rm u}$ is given by Eq.~(\ref{dun}).
As a result, Eq.~(\ref{eff}) becomes
\begin{equation}
{\cal L}_{\rm eff}=-2^{1/4}G_F^{1/2}m_b\overline bbH(1+\delta_{\rm u})
(1+\delta_{\rm nu}).
\end{equation}

\section{Results}

In the previous section, we have constructed a low-$M_H$ effective Lagrangian
for the $b\bar bH$ interaction in the Standard Model by integrating out the
$t$ quark.
Using this Lagrangian, we can now compute $\Gamma\left(H\to b\bar b\right)$
including the ${\cal O}(G_Fm_t^2)$ and ${\cal O}(\alpha_sG_Fm_t^2)$
corrections.
By accommodating also the strong, electromagnetic, and residual weak
corrections, we obtain the full Standard-Model prediction,
which we may write in a factorized form,
\begin{equation}
\label{gam}
\Gamma\left(H\to b\bar b\right)=\Gamma_{\rm QCD}\left(H\to b\bar b\right)
(1+\delta_{\rm u})^2(1+\delta_{\rm nu})^2(1+\Delta_{\rm weak}-x_t)
\left(1+{\alpha\over\pi}Q_b^2\delta_{\rm QED}\right),
\end{equation}
where $\delta_{\rm u}$ and $\delta_{\rm nu}$ are listed in
Eqs.~(\ref{dun},\ref{dnu}), respectively, and $\Delta_{\rm weak}$ and
$\delta_{\rm QED}$ may be found in Ref.~\cite{hff}.
A low-$M_H$ approximation for $\Delta_{\rm weak}$ is given by Eq.~(\ref{weak}).
$\Gamma_{\rm QCD}\left(H\to b\bar b\right)$ contains the tree-level result of
Eq.~(\ref{born}) along with its QCD corrections,
which we have reviewed in the Introduction.
Adopting the on-shell definition of quark mass, the ${\cal O}(\alpha_s)$
result reads
\begin{equation}
\Gamma_{\rm QCD}\left(H\to b\bar b\right)=
\Gamma_0\left(H\to b\bar b\right)\left(1+C_F{\alpha_s\over\pi}
\delta_{\rm QED}\right).
\end{equation}
For $m_b\ll M_H/2$, $\delta_{\rm QED}$ may be expanded as \cite{hff}
\begin{equation}
\delta_{\rm QED}=-{3\over2}\ln{M_H^2\over m_b^2}+{9\over4}+
{\cal O}\left({m_b^2\over M_H^2}\ln{M_H^2\over m_b^2}\right).
\end{equation}

It is interesting to study how the $m_t$-dependent term in Eq.~(\ref{weak})
is affected by QCD corrections.
Toward this end, we consider the product
\begin{eqnarray}
\label{fin}
&\n\n&\left(1+C_F{\alpha_s\over\pi}\delta_{\rm QED}\right)(1+\delta_{\rm u})^2
(1+\delta_{\rm nu})^2
\nonumber\\
&\n\n&
=1+{3\over2}C_F{\alpha_s\over\pi}\left(-\ln{M_H^2\over m_b^2}+{3\over2}\right)
+x_t\left[1-3C_F{\alpha_s\over\pi}\left({1\over2}\ln{M_H^2\over m_b^2}+\zeta(2)
+{1\over4}\right)\right]+\cdots,
\end{eqnarray}
where the ellipsis represents terms of
${\cal O}(\alpha_sm_b^2/M_H^2\ln(M_H^2/m_b^2))$, ${\cal O}(\alpha_s^2)$,
and ${\cal O}(G_F^2m_t^4)$.
We recover the notion that, in Electroweak Physics, one-loop
${\cal O}(G_Fm_t^2)$ terms get screened by their QCD corrections.
In the present case, the screening effect is extraordinarily strong.
In fact, for $M_H=60$~GeV ($2M_W$), the ${\cal O}(\alpha_sG_Fm_t^2)$ correction
compensates 71\% (75\%) of the ${\cal O}(G_Fm_t^2)$ term.
Here, we have employed $m_b=4.72$~GeV \cite{dom} and evaluated $\alpha_s(\mu)$
at renormalization scale $\mu=M_H$.
As a normalization point, we have used $\alpha_s(M_Z)=0.118$ \cite{bet}.
However, we should bear in mind that the ${\cal O}(G_Fm_t^2)$ term is
incidentally small due to the almost complete cancellation of the universal and
non-universal contributions.

The presence of large logarithmic terms like $\alpha_s\ln(M_H^2/m_b^2)$ is,
of course, detrimental for the speed of convergence of the QCD perturbation
series.
However, from the organization of Eq.~(\ref{gam}) it is evident that these
logarithms may be rendered harmless in the usual way, by introducing the
running $b$-quark mass of the $\overline{\rm MS}$ scheme evaluated at scale
$M_H$.
The appropriate formula for $\Gamma_{\rm QCD}\left(H\to b\bar b\right)$
is listed in Eq.~(21) of Ref.~\cite{gkw}.
By expanding the weak correction factor in Eq.~(\ref{gam}),
$(1+\delta_{\rm u})^2(1+\delta_{\rm nu})^2(1+\Delta_{\rm weak}-x_t)$,
we recover Eq.~(\ref{weak}) along with the QCD correction factor that
multiplies the $m_t$-dependent term of Eq.~(\ref{weak}).
The latter reads
\begin{eqnarray}
1-3(\zeta(2)+1)C_F{\alpha_s\over\pi}&\n=\n&
1-2\left({\pi\over3}+{2\over\pi}\right)\alpha_s\nonumber\\
&\n\approx\n&1-3.368\,\alpha_s.
\end{eqnarray}
In this way of presenting our result, the screening effect amounts to
$-42\%$ ($-37\%$) for $M_H=60$~GeV ($2M_W$).

At this point, we should compare our analysis of
$\Gamma\left(H\to b\bar b\right)$ with the result of a recent work \cite{kwi}.
According to Eq.~(12) in Ref.~\cite{kwi}, the ${\cal O}(G_Fm_t^2)$ term gets
dressed by the factor
\begin{equation}
\label{kfa}
K=1+{\alpha_s\over\pi}\left(-22-{2\over3}\pi^2+12\ln{M_H^2\over m_b^2}\right),
\end{equation}
which has to be contrasted with the square bracket in our Eq.~(\ref{fin}).
We observe that the expression in Eq.~(\ref{kfa}) is significantly larger than
our $K$ factor, the ratio of the two being 8.0 (11.8) at $M_H=60$~GeV ($2M_W$).
In other words, the authors of Ref.~\cite{kwi} find an enhancement of the
${\cal O}(G_Fm_t^2)$ term by 130\% (194\%),
whereas we find a reduction by 71\% (75\%).
Comparing Eqs.~(\ref{fin},\ref{kfa}), we can trace the source of this
discrepancy.
In fact, up to terms of ${\cal O}(\alpha_s^2G_Fm_t^2)$, we can rewrite
Eq.~(\ref{kfa}) as
$Kx_t=2\delta_{\rm u}+2\delta_{\rm nu}[1+C_F(\alpha_s/\pi)\delta_{\rm QED}]$,
i.e., the interference of the
${\cal O}(G_Fm_t^2)$ universal term and the ${\cal O}(\alpha_s)$ term,
$2\delta_{\rm u}C_F(\alpha_s/\pi)\delta_{\rm QED}$, is missing in
Eq.~(\ref{kfa}).

\section{Conclusions}

We computed the two-loop ${\cal O}(\alpha_sG_Fm_t^2)$ non-universal correction
to $\Gamma\left(H\to b\bar b\right)$, which arises from the $b\bar bH$ vertex
and the $b$-quark wave function, by means of a low-energy theorem.
Combining this result with the universal correction in the same order,
which is due to the renormalizations of the Higgs-boson wave function and
vacuum expectation value and contributes to any Higgs-boson decay to fermions
or intermediate bosons,
we obtained the full ${\cal O}(\alpha_sG_Fm_t^2)$ correction to
$\Gamma\left(H\to b\bar b\right)$.
This correction screens the positive ${\cal O}(G_Fm_t^2)$ term by 71\% to 75\%
for $M_H$ between 60~GeV and $2M_W$.
As a consequence, the sensitivity of $\Gamma\left(H\to b\bar b\right)$ to the
top quark, which, at one loop, is already seven times weaker than in the case
of the other fermionic decay modes, is practically quenched.
We presented a master formula for $\Gamma\left(H\to b\bar b\right)$, which
makes full use of the present knowledge of radiative corrections to this
quantity.
We conclude that the residual theoretical uncertainty due to unknown
higher-order corrections is likely to be negligible as compared to the
envisaged experimental error.

\bigskip
\centerline{\bf ACKNOWLEDGMENTS}
\smallskip\noindent
We are grateful to Bodo Lampe for beneficial discussions.

\begin{figure}[p]

\vspace*{-15.0cm}
\hspace*{-0.5cm}
\begin{turn}{90}%
\epsfxsize=25cm \epsfbox{hbb.fig1}
\end{turn}
\vspace*{-5.5cm}

\centerline{\bf (a)}

\vspace*{-11.0cm}
\hspace*{-5.0cm}
\begin{turn}{90}%
\epsfxsize=18cm \epsfbox{hbb.fig2}
\end{turn}
\vspace*{-2.7cm}

\centerline{\bf (b)}
\vspace*{1.0cm}

\centerline{\bf FIGURE CAPTION}

\caption{\label{one}Feynman diagrams pertinent to the $b$-quark self-energy
in (a) ${\cal O}(G_Fm_t^2)$ and (b) ${\cal O}(\alpha_sG_Fm_t^2)$.}

\end{figure}

\end{document}